\documentclass{article}[12pt]
\usepackage{amsmath}
\usepackage{amssymb}
\usepackage{amsthm}
\usepackage{amsfonts}
\usepackage[dvips]{graphicx}
\usepackage{color}

\def\abstract#1{\vskip 7mm
        \begin{center}{\large Abstract}\par \smallskip
                \begin{minipage}[c]{12cm}
                        \small #1
                \end{minipage}
        \end{center}
}

\def\title#1{\begin{center}{\Large\bf #1}\end{center}}
\def\author#1{\vskip 5mm \begin{center}{#1}\end{center}}
\def\address#1{\begin{center}{\it #1}\end{center}}

\newcommand{\ssmatrix}[4]%
{\begin{pmatrix} #1 & #2 \\ #3 & #4 \end{pmatrix}}
\makeatletter
\@ifundefined{lesssim}{}{}
\@ifundefined{gtrsim}{}{}
\def\vereq#1#2{\lower3pt\vbox{\baselineskip1.5pt \lineskip1.5pt
\ialign{$\m@th#1\hfill##\hfil$\crcr#2\crcr\sim\crcr}}}
\makeatother

\newtheorem{lemma}{Lemma}
\newtheorem{proposition}{Proposition}

    {\mbox{}\hfill\mbox{\rm\normalsize$\Box$}\\%
    }
\usepackage{color}
\input{colordvi.tex}

\begin{document}

\title{Dynamical degrees of freedom for higher genus Riemann surface in (2+1)-dimensional general relativity}

\author{Masaru Siino}

\address{Department of Physics, Tokyo Institute of Technology, Tokyo 152-8551, Japan}

\abstract{A homogeneous two-dimensional metric including the degrees of freedom of Teichm\"{u}ller deformation is developed. The Teichm\"{u}ller deformation is incorporated by affine stretching of complex structure. According to Yamada's investigation by pinching parameter, concrete formulation for a higher genus Riemann surface can be realized. We will have a homogeneous standard metric including the dynamical degrees of freedom as Teichm\"{u}ller deformation  in a leading order of the pinching parameter, which would be treated as homogeneous anisotropic metric for a double torus universe, which satisfy momentum constraints.}

\section{Introduction}
In spite of the absence of propagating gravitational wave, the gravitation in (2+1)-dimensional spacetime has been investigated from the various viewpoints\cite{3D,3DBH,HNM,SF,HN2}, and its importance is still now surviving supported by the new concepts\cite{AC,MW,KS}.
Probably, that will be recognized as the result of that there still remain global degrees of freedom of the spacetime in (2+1)-dimensional gravity.

As pointed out by Hosoya, Nakao and Moncrief\cite{HNM}, it seems clear that the two-dimensional spatial geometry
is understood as a Riemann surface with complex structure in York's timeslicing\cite{JY}.
In other words, in (2+1)-dimensional spacetime, a spacelike hypersurface is locally conformally flat, and then the `geometry' is represented by the conformal class of itself. That will be corresponding to the complex structure, in the
identification between the conformal transformation and bi-holomorphic transformation.
To investigate such a complex structure on a Riemann surface one may conventionally prefer to a local complex coordinate $(z,\bar{z}), z=x+i y$. 
The Riemann surface is a two-dimensional surface with complex structure,
which is with a set of local complex coordinate, recognizing an equivalence relation by bi-holomorphic function.
In a sense of Riemannian geometry, the complex structure is just an equivalence class
of Riemannian two-geometry by a conformal mapping.
We would realize the degrees of freedom of the complex structure by the Teichm\"{u}ller
deformations parametrized by parameters in Teichm\"{u}ller space\cite{FK,IT}. 

In \cite{HNM}, it was shown that the transverse traceless part of extrinsic curvature $\widetilde{K}_{ab}$ is a holomorphic quadratic differential spanning a complex vector space of ($3g - 3$) dimensions and  one dimension exceptionally  for a torus.
Determining the holomorphic quadratic differential for a torus universe to solve the momentum constraint, the Hamiltonian constraint would be resolved for pure gravity on homogeneous spacelike hypersurface in York's gauge choice.
Moreover, these aspects can be reformulated\cite{SF}, considering a standard homogeneous metric containing the global degrees of freedom as Teichm\"{u}ller parameters of torus  in the context of affine stretching\cite{IT}.
When we consider a homogeneous spatial hypersurface by the definition of time coordinate, the Einstein equations will become an ordinary differential equation about time of the expansion coefficients of the Teichm\"{u}ller deformation by the holomorphic quadratic differential.
For a higher genus $g>1$ Riemann surface, however, similar analysis is not straightforward, and the holomorphic quadratic differential accounts for its difficulty. There, that is not a constant and also is not given in concrete expression, remarkably containing $4g-4$ zeros. So, it seems difficult to get a homogeneous standard metric such that the Einstein equation is reduced to ordinary differential equations of time.


In the present study, we attempt to treat the complex structures by the expansions around the Riemann surface with highly pinching narrow bridges. Yamada\cite{YA} have established the description of such an expanded complex structure in full order of the pinching parameter $\epsilon$.
We will have concrete expression of holomorphic quadratic differential in the expansion. Especially in leading order of $\epsilon$, we will determine the homogeneous standard metric to study the pure gravitational dynamics of a double torus universe.

First of all we give a formal definition of the Teichm\"{u}ller deformation by holomorphic quadratic differential in the context of the affine stretching in the second section. The leading order of  holomorphic quadratic differential by the pinching parameter is developed for a double torus according to Yamada's work in the third section. The fourth section provides a homogeneous standard metric of it and the final section is devoted for summary and discussions.

\section{Affine stretching}
Our purpose of the present study is to formulate the geometrical degrees of freedom in (2+1)-dimensional gravity where spatial section of the spacetime realized as a higher genus Riemann surface.
From the view point of metric tensor, the Riemann surface is the conformal class of the Riemannian geometry on two dimensional manifold which is a spacelike section $\Sigma$ of the spacetime because complex local coordinate $\{z_{\alpha}\}$ clarify that conformal mapping is identified to bi-holomorphic function. 

As in an appropriate gauge choice,  the time covariance would be recognized as in the conformal degrees of freedom\cite{SF,HN2} and then the geometrical degrees of freedom will be identified to those of the Riemann surface. Nevertheless, since two-dimensional Riemannian geometry admitting a locally conformally flat coordinate, the local degrees of freedom are absence. (That implies there is no gravitational wave.)
We will account for the global degrees of freedom, which is regarded as Teichm\"{u}ller deformation\cite{IT}.

The Teichm\"{u}ller deformation will be realized as a quasiconformal mapping of the complex structure $(R,\{z_{\alpha}\})$ of the Riemann surface $R$.
Then the degrees of freedom is represented by the Beltrami coefficient $\mu=f_{,\bar{z}}/f_{,z}$ of the
quasiconformal mapping $w=f(z,\bar{z})$ since the pull back of the local metric function $|dw|^2$ is given by $|f_{,z}|^2|dz+\mu d\bar{z}|^2$, and we are required to solve Beltrami equation.
With vanishing $\mu$, $f$ is a bi-holomorphic function ($\sim$ conformal mapping).
One may call $\mu$ complex dilatation which represents ellipticity and rotation of 
the infinitesimal circular image.

Another way to realize the degrees of freedom for Teichm\"{u}ller deformation
is to consider affine transformation of the local complex coordinate\cite{IT}. In the following, we see a two dimensional metric including the degrees of freedom of the Teichm\"{u}ller deformation is given as a result of affine stretching along the coordinate incorporated by the holomorphic quadratic differential.
The affine transformation of the local coordinate $z\mapsto w(z,\bar{z})=\alpha z+\beta \bar{z}, (\alpha,\beta \in \mathbb{C},|\alpha|>|\beta|)$ reduces to the 
\begin{align}
z \mapsto z +k \bar{z}
\label{eqn:as}
\end{align}
by the redefinition of the local coordinate $z, w$
and assign the ellipticity by $0<k = \beta / \alpha<1$, which will be real by the rotation of the complex coordinates and fixed.

Of course, the local coordinate can be extended by analytic continuation of bi-holomorphic mapping. 
Then the pull back of the Euclidean metric $|dw|^2$ by the bi-holomorphic mapping $z=h(\zeta)$ is given by

\begin{align}
|dw|^2=|h'|^2|d\zeta+k\frac{\bar{h'}^2}{|h'^2|}d\bar{\zeta}|^2.
\label{eqn:asm}
\end{align}

Here $\varphi\equiv(h')^2$ is regarded as a holomorphic quadratic differential since it is transformed by the change of local coordinate $(U_j,z_j)\mapsto (U_k,z_k)$ as $\varphi_k(z_k)=\varphi_j(z_j)(dz_j/dz_k)^2$.
Therefore the quasiconformal mapping is determined by a holomorphic quadratic differential $\varphi$ and $k$.

It is well known from the Riemann Roch Theorem\cite{FK}, that the holomorphic quadratic differential spans a complex vector space $A_2(R)$ whose dimension is $3g-3$.  
In the context of Riemannian geometry the space of complex structure is identified to conformal class of Riemannian geometry on the surface $\Sigma$, and $A_2(R)$ is homeomorphic to Teichm\"{u}ller space $T(R)$ (the space of Teichm\"{u}ller deformation or quasi conformal mapping)
\[
T(R)=\frac{\cal M}{\{\rm Conf(\Sigma)\} \{\rm Diffeo(\Sigma)\}},
\]
where $\cal M$ is the space of Riemannian geometry on $\Sigma$ to become quotient by the conformal equivalence and diffeomorphism\footnote{Here we do not consider modular transformation. On considering quantum theory, we should redefine the modular space as the modular class of the Teichm\"{u}ller space\cite{HN2}.}.
Then considering a quasiconformal mapping $f$ which is affine transformation in a local coordinate $z$ as (\ref{eqn:as}), its Beltrami coefficient is represented by an element $\varphi$ of $A_2(R)$ as
\[
\mu_f=k\frac{\bar{\varphi}}{|\varphi|},
\]
which is a quasiconformal mapping as Teichm\"{u}ller deformation for $(k,\varphi)$\footnote{$k$ is an fixed parameter, which will have any technical role in formulation of canonical variables in our forthcoming work\cite{FW}}.

\begin{figure}[hbtp]
\centering
\includegraphics[width=7cm,clip]{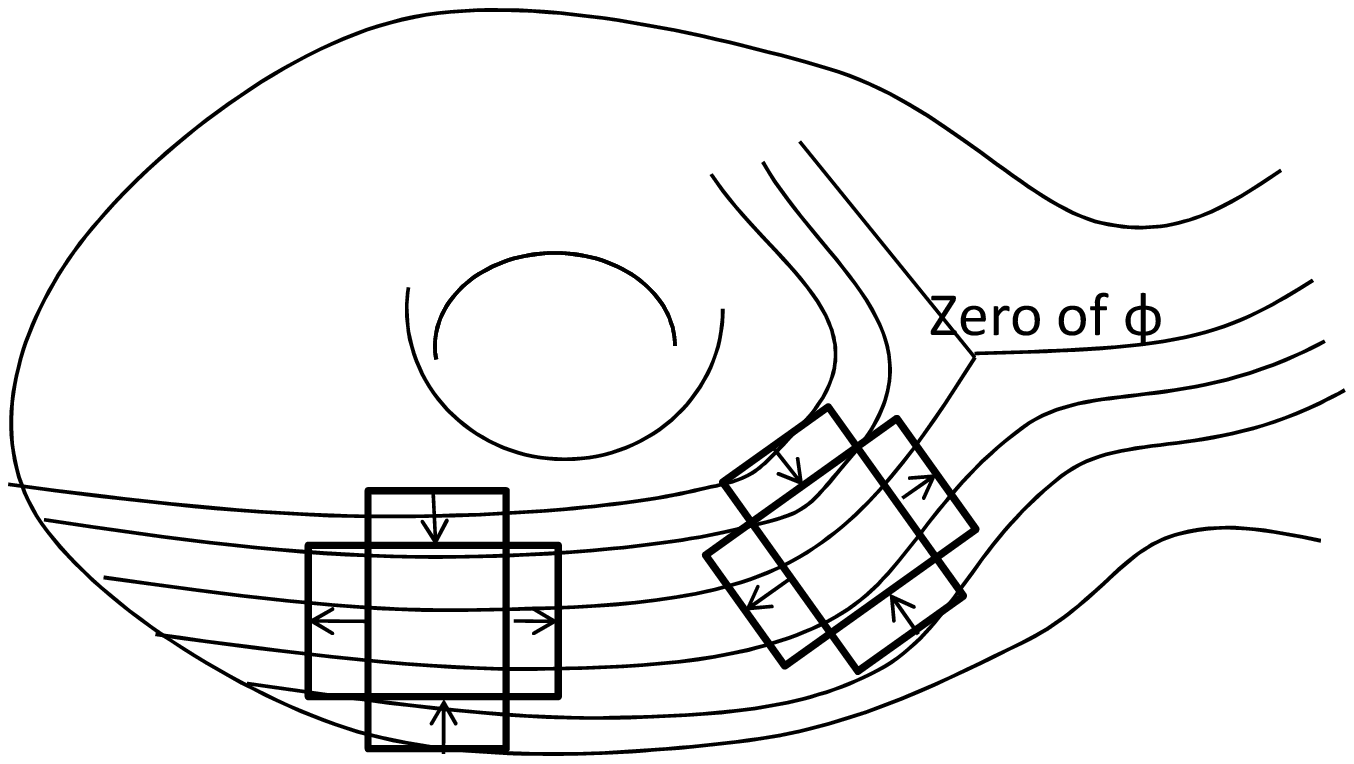}
\caption{A horizontal foliation which is introduced by the root of a holomorphic quadratic differential $\varphi$ is illustrated. That indicates the direction of the affine stretching by  $\varphi$. }
\label{fig:horizon}
\end{figure}

These aspects roughly illustrated in Figure \ref{fig:horizon}. Since $\varphi d\zeta^2$ is holomorphic quadratic differential, its root ($\sim h' d\zeta$) gives holomorphic one-form $dz$. Then the integration   of it $z=\int h'd\zeta$ gives not a global coordinate but  horizontal (or vertical) foliation \cite{FG} by analytic continuation,
since such a local flat coordinate cannot be extended to global one beyond the zeros of $\varphi$ as suggested by the uniformization theorem.
In other words, the horizontal foliation has branches at the zeros as illustrated.
Then the local Euclidean metric of local coordinate $(z,\bar{z})$ is affine stretched by (\ref{eqn:as}). 

The equivalence class of this affine stretched complex structure is given by the conformal
 class of the metric (\ref{eqn:asm}).
Usually we will choose a conformally invariant one $\tilde{g}_{ab}:=g_{ab}/\sqrt{g}$ as a representative of the conformal class of the affine stretching:
\begin{align}
|\widetilde{dw}|^2=|d\zeta+k\frac{\bar{\varphi}}{|\varphi|}d\bar{\zeta}|^2.
\label{eqn:rep}
\end{align}
Consequently, the Teichm\"{u}ller deformation generating the Teichm\"{u}ller space $T(R)$.
(For the case of torus, exceptionally the dimension of $A_2(R)$ is one and
since the $\mu_f$ is essentially constant, the global affine transformation
uniformly gives a metric function $|dw|^2$ written by the well known Teichm\"{u}ller  parameter $\tau$ for a torus\cite{SF}.)

To have the holomorphic quadratic differential, we consider holomorphic one-form firstly.
For a Riemann surface $R$ of genus $g$ whose canonical homology basis are given by $(A_1,...,A_g,B_1,...,B_g)$, there exists $g$ holomorphic one-forms $\nu_i$ normalized as
\begin{align}
\int_{A_i}\nu_j= \delta_{ij}.
\end{align}
These forms can be neatly encapsulated in a unique singular bilinear two-form\cite{YA}
\begin{align}
\omega(u,v)=\{\frac1{(u-v)^2}+{\rm regular\ \ terms }\}du dv\label{eqn:ndsk1}
\end{align}
with normalization
\begin{align}
\int_{A_i} \omega(u,\cdot)=0 \label{eqn:ndsk2}
\end{align}
and this is a differential of second kind, where $u,v$ is complex local coordinates.

From the Riemann bi-linear relations, we see 
\begin{align}
\nu_i(u)=\frac1{2\pi i}\int_{B_i}\omega(u,\cdot).
\end{align}

Then the quadratic differential spans a complex vector space,
\begin{align}
A_2(R)=\langle\varphi_{ij}\rangle_{\mathbb C}=\langle\nu_i\cdot\nu_j |i,j=1,...,g\rangle_{\mathbb C} .
\end{align}
For a higher genus $(g>1)$ Riemann surface, this is $3g-3$ dimensional from the Riemann Roch theorem.
From the Teichm\"{u}ller's theorem\cite{IT}, the Teichm\"{u}ller space $T(R)$ is homeomorphic to $A_2(R)$\footnote{To be exact, $A_2(R)_1=\{\varphi \in A_2(R) | ||\varphi ||_1<1\}$ ($||\varphi ||_1:=2\int\int_R|\varphi(z)|dx dy$) is homeomorphic to $T(R)$.} and furthermore to ${\bf R}^{6g-6}$.

As an example, we will consider the case of a torus. 
So, different from the higher genus, there is only one independent holomorphic one-form $\nu=d\zeta$, which is constant.
Then the quadratic form $(h')^2d\zeta^2$ is given by two real constant parameters as $(\varphi_R+i\varphi_I)d \zeta^2$.
In other words, 
\[h'd\zeta=(\nu_R+i\nu_I)d\zeta
\]
is complex and constant.
Consequently the affine stretched metric (\ref{eqn:asm}) is given by
\begin{align}
|dw|^2&=|(\nu_R+i\nu_I)d\zeta+(\nu_R-i\nu_I)k d\bar{\zeta}|^2\\
&=(d\zeta_x,d\zeta_y)
\left(\begin{array}{cc}
(1+k)\nu_R^2+(1-k)\nu_I^2 & -2k\nu_R\nu_I \\
-2 k\nu_R\nu_I & (1+k)\nu_I^2+(1-k)\nu_R^2 \end{array}\right)
\left(\begin{array}{c}
d\zeta_x \\
d\zeta_y \end{array}\right)\label{eqn:mtx}.
\end{align}

Introducing a conventional Teichm\"{u}ller parameters, instead of $(\nu_R,\nu_I)$
\[
\xi=\frac{-2k\nu_R\nu_I}{(1-k)\nu_R^2+(1+k)\nu_I^2},\ \  \eta=\frac{\sqrt{1-k^2}(\nu_R^2+\nu_I^2)}{(1-k)\nu_R^2+(1+k)\nu_I^2},
\]
the conformally invariant representative $\tilde{g}_{ab}=g_{ab}/\sqrt{g}$ gives well known standard metric for torus\cite{SF,IT};
\begin{align}
\frac1{\eta}.\left(\begin{array}{cc}
\xi^2+\eta^2 & \xi \\
\xi &1 \end{array}\right).
\end{align}
For vanishing $k$, $|dw|^2$ reduces to the conformally Euclidean form $|\varphi||d\zeta|^2$ meaning equivalent complex structure to the original one.

\section{pinching parameter and complex structure}
It may be a consensus that it is strongly difficult to give a concrete expression of the differential forms $\omega(u,v), \nu_i, \varphi_{ij}$ for a general Riemann surface with higher genus $g>1$.

On the other hand, for a genus $g$ Riemann surface including a narrow bridge structure, 
one can approximately compose the differentials $\omega(u,v), \nu_i, \varphi_{ij}$
from the differentials on two Riemann surfaces with lower genus,
from which the genus $g$ Riemann surface is obtained by sewing.
Therefore, we only consider the case where a Riemann surface can be 
thought of as one composed 
of two Riemann surfaces connected by a narrow bridge. 
A general method for calculating $\omega(u,v)$ for any two sewn Riemann 
surfaces has been given by Yamada~\cite{YA}.

\begin{figure}[htpb]
\centering
\includegraphics[width=9cm]{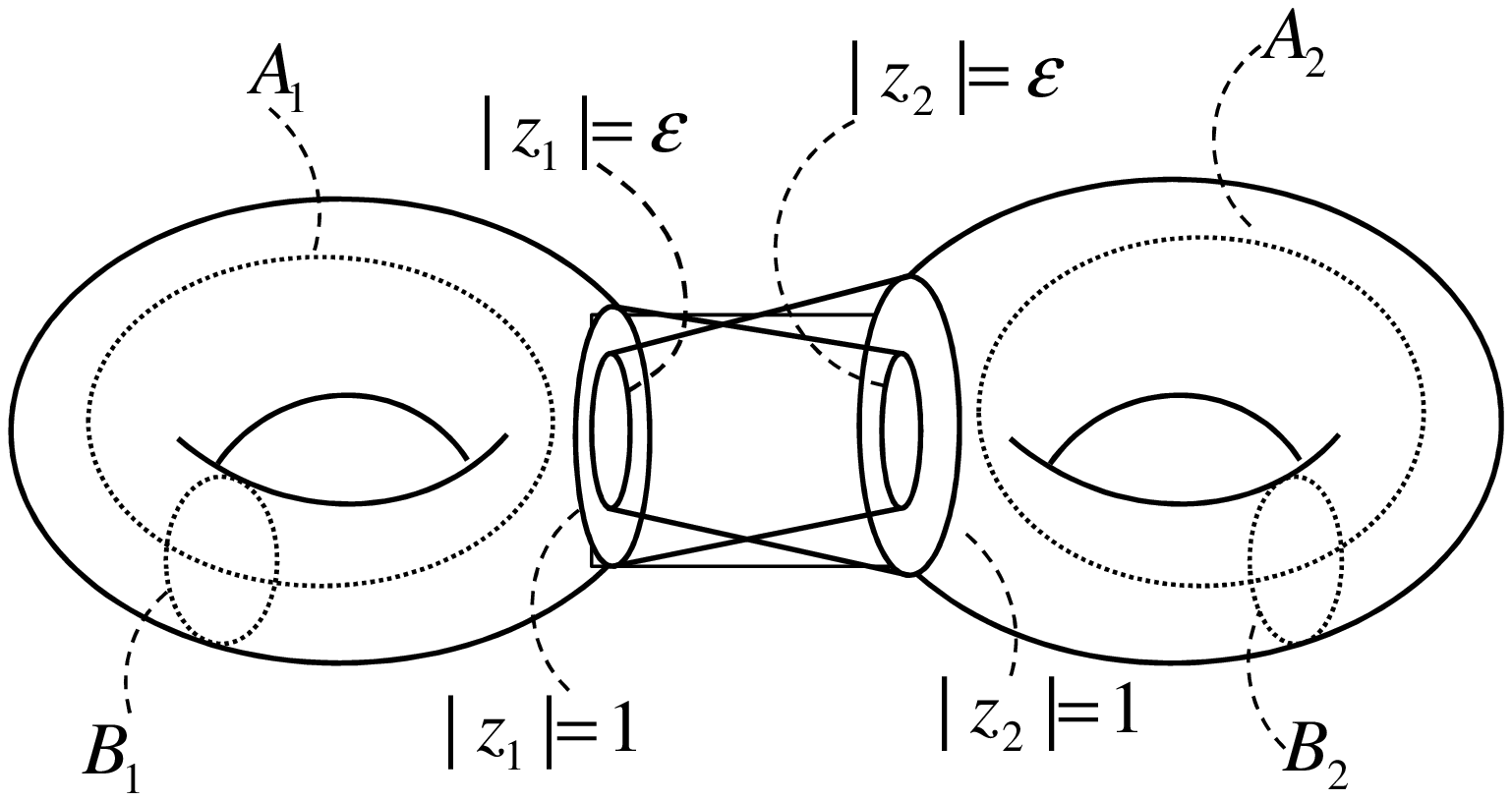}
\caption{Two tori are sewn together by identifying the annular regions 
$|\epsilon|\leq |z_a|\leq 1$ via the relation
$z_1 z_2 =\epsilon$. }
\label{fig:two-tori}
\end{figure}

In general, two compact Riemann surfaces $S_1$ and $S_2$ of genus $g_1$ and $g_2$
can be sewn together, giving a Riemann surface of genus $g_1+g_2$.
Choose complex local coordinates $z_a$ on $S_a$ ($a=1,2$), 
and excise the two disks $|z_a|<|\epsilon|$, 
where $\epsilon$ is a complex parameter satisfying $|\epsilon|<1$.
The centers of disks are taken at a point $z_a =0$.
The two surfaces are sewn together by identifying the annular regions 
$|\epsilon|\leq |z_a|\leq 1$ via the relation
\[z_1 z_2 =\epsilon.\]

By the way, for a pinching parameter $\epsilon$,
${\cal A}_a=\{\epsilon\leq |z_a|\leq 1\}\subset S_a$ is a cylinder to be identified.
Let ${\cal C}_a(z_a)\subset {\cal A}_a$ denote a simple closed, anti-clockwise
oriented contour parametrized by $z_a$ surrounding the puncture at $z_a=0$,
which is in relation of sewing ${\cal C}_1(z_1)\sim -{\cal C}_2(z_2)$
via $z_1z_2=\epsilon$. 
In the following, we refer and adopt following three results of \cite{YA}, without proof.

Yamada\cite{YA} had started from the following lemma.
\begin{lemma}
\begin{align}
\omega(u,v)=\omega^{(a)}(u,v)\delta_{ab}+\frac1{2\pi i}\oint_{{\cal C}_a(z_a)}\left(\omega(v,z_a)\int_0^{z_a}\omega^{(a)}(u,\cdot)\right)
\end{align}
for $u\in S_a, v\in S_b  (a,b=1,2)$.
\end{lemma}
$\omega^{(a)}$ is a unique bilinear two-form for a Riemann surface $S_a$.

Then it was demonstrated that the following expression expanded by the power of $\epsilon$ will be possible.
\begin{lemma}
\begin{align}
\omega(u,v)=
\begin{cases}
\omega^{(a)}(u,v)+a_a(u)X_{\bar{a}\bar{a}}a_a^T(v) & u,v\in S_a \\
a_a(u)(-I+X_{\bar{a}a})a_{\bar{a}}^T(v) & u\in S_a, v\in S_{\bar{a}} \\
\end{cases}
\end{align}
where $a_a(u)[k]$ is row vector
and $X_{ab}[k,l]$ is matrix indexed by $k,l=1,2,...$,
defined by
\begin{align}
X_{ab}[k,l]&=\frac{\epsilon^{(k+l)/2}}{\sqrt{kl}}\frac1{(2\pi i)^2}\oint_{{\cal C}_a(u)}\oint_{{\cal C}_b(v)}u^{-k}v^{-l}\omega(u,v). \\
a_a(u)[k]&=\frac{\epsilon^{k/2}}{2\pi i\sqrt{k}}\oint_{{\cal C}_a(z_a)}z_a^{-k}\omega^{(a)}(u,z_a),\label{eqn:a}
\end{align}
where $\bar{a}$ means the complement of $a$.
\end{lemma}

Besides, $X_{ab}$ is given in terms of 
\begin{align}
A_a[k,l]=\frac{\epsilon^{k/2}}{2\pi i\sqrt{k}}\oint_{{\cal C}_a(u)}u^{-k}a_a(u)[l] .\label{eqn:A}
\end{align}

The consistency is guaranteed by this proposition.
\begin{proposition}
\begin{align}
X_{aa}=A_a(I-A_{\bar{a}}A_a)^{-1} \\
X_{a\bar{a}}=I-(I-A_a A_{\bar{a}})^{-1} 
\end{align}
Here, 
\begin{align}
(I-A_a A_{\bar{a}})^{-1}=\sum_{n\geq 0}(A_a A_{\bar{a}})^n
\end{align}
and is convergent as a power series in $\epsilon$ for $|\epsilon|<1$.
\end{proposition}

Though the full expression is complicated, the leading order of contribution of the pinching parameter $\epsilon$ can be simply determined.
Now we calculate the leading contribution of $\epsilon$ from the lemma's and the proposition for a $g=2$ Riemann surface (double torus) with the pinching parameter. $S_1$ and $S_2$ are tori and their Teichm\"{u}ller parameters are $\tau_1$ and $\tau_2$, respectively.

First of all, we remind the definitions of the Weierstrauss's zeta function $\zeta(z)$ and elliptic function ${\cal P}(z)$\cite{FK}, for a torus with a Teichm\"{u}ller parameter $\tau$; 
\begin{align}
\zeta(z)=\frac1{z}+\sum_{m,n\in {\mathbb Z}, m^2+n^2\neq 0}\left[ \frac1{z-m-n\tau}+\frac1{m+n\tau}+\frac{z}{(m+n\tau)^2}\right],
\end{align}
\begin{align}
{\cal P}(z)=\frac1{z^2}+\sum_{m,n\in {\mathbb Z}, m^2+n^2\neq 0}\left[ \frac1{(z-m-n\tau)^2}-\frac1{(m+n\tau)^2}\right].
\label{eqn:wp}
\end{align}
In (\ref{eqn:wp}) the last term of summation $\sum (m+n\tau)^{-2}=E_2(\tau)$ is convergent and known as one of the Eisenstein series $E_{2k}=\sum \frac1{(m+n\tau)^{2k}}$, such that
the zeta function will be expanded as $\zeta(z)=\frac1{z}-\sum_{k=1}^{\infty}E_{2k+2}z^{2k+1}$. 
From ${\cal P}(z)=-\frac{d}{dz}\zeta(z)$, also ${\cal P}(z)$ can be expanded as ${\cal P}(z)=\frac1{z^2}+\sum_{k=1}^{\infty}(2k+1) E_{2k+2}z^{2k}$.

By definition, it is obvious that ${\cal P}(z)$ is periodic function as
\begin{align}
{\cal P}(z)={\cal P}(z+m+n\tau), \ \ m,n\in \mathbb Z.
\end{align}
Integrating it, one may realize that $\zeta(z)$ is with quasi-periodic properties
\begin{align}
\zeta(z)=\zeta(z+m+n\tau)-(\eta_1 m+\eta_{\tau} n)
 \ \ m,n\in \mathbb Z.
\end{align}
With the aid of Legendre's equation $\eta_1 \tau-\eta_{\tau}=2\pi i$,
we conclude that $\eta_1=E_2$ and $\eta_\tau=\tau E_2-2\pi i$.
Defining $P_1(\tau,z)=\zeta(z)-z E_2(\tau)$, the periodicity is simply expressed as $P_1(\tau,z)=P_1(\tau,z+m+\tau n)+2\pi i n, \ m,n\in\mathbb{Z}$.

From (\ref{eqn:ndsk1}), ${\cal P}(z)$ will provide  a unique bilinear two-form (which is differential of second kind) for a torus parametrized in the usual way but not normalized.
We see
\begin{align}
\omega^{(a)}(u,v)=({\cal P}(\tau_a,u-v)+E_2(\tau_a)) dudv =:P_2(\tau_a,u-v)dudv \ \ \ (a=1,2),
\label{eqn:b2f}
\end{align}
satisfies (\ref{eqn:ndsk2}) by the quasi-periodicity of $P_1(z)$, (c.f. $P_2=-\frac{d}{dz}P_1$), where $\tau_a$ is the Teichm\"{u}ller parameter of each torus ($\sim S_a$).

Now we are back to genus 2 Riemann surface.
From (\ref{eqn:a}) and (\ref{eqn:A}), one may have contributions of leading order of $\epsilon$,
\begin{align}
a_a(u)[1]&=\frac{\epsilon^{1/2}}{2\pi i}\oint_{{\cal C}_a(z_a)}z_a^{-1}\omega^{(a)}(u,z_a)
=\epsilon^{1/2}P_2(\tau_a,u)du \\
A_a[1,1]&=\frac{\epsilon^{1/2}}{2\pi i}\oint_{{\cal C}_a(u)} u^{-1}a_a(u)[1] 
=\epsilon E_2(\tau).
\end{align}

Then a basis of our analysis is given;
\begin{align}
\omega(u,v)=\begin{cases}P_2(\tau_a,u-v)dudv+\epsilon^2 E_2(\tau_{\bar{a}})P_2(\tau_a,u)P_2(\tau_a,v)du dv, \ \ (u\in S_a, \ v\in S_a)\\
-\epsilon P_2(\tau_a,u) P_2(\tau_{\bar{a}},v)du dv, \ \  (u\in S_a, \ v\in S_{\bar{a}})
\end{cases}.
\end{align}

Deductively, we see the holomorphic one-form for the double torus with pinching parameter $\epsilon$,
\begin{align}
\nu_1(u\in S_1)&= \frac1{2\pi i}\int_{B_1(v)} \left[P_2(\tau_1,u-v)+\epsilon^2 E_2(\tau_1)P_2(\tau_1,u)P_2(\tau_1,v)\right]du dv \\
&=-\frac1{2\pi i}\left[ P_1(\tau_1,u-v+\tau_1)-P_1(\tau_1,u-v)\right]du\\
&-\frac1{2\pi i}\epsilon^2 E_2(\tau_2)P_2(\tau_1,u)\left[ P_1(\tau_1,v+\tau_1)-P_1(\tau_1,v)\right]du \\
&=du+ \epsilon^2 E_2(\tau_2)P_2(\tau_1,u)du\\
\nu_1(u\in S_2)& =-\frac1{2\pi i}\int_{B_1(v)}\epsilon P_2(\tau_2,u)P_2(\tau_1,v)du dv \\
&=-\epsilon P_2(\tau_2,u) du.
\end{align}
As expected, they coincides to that of torus with vanishing $\epsilon$.
Then we can identify its complex structure from them. 
For example\cite{IT}, the holomorphic one-form determines a period matrix by
\begin{align}
&\int_{A_j}\nu_i=\delta_{ij} \\
\Omega_{ij}&=\int_{B_j}\nu_i\\
&=\left(
\begin{array}{c|c}
\int_{B_1}\nu_1=\tau_1+2\pi i\epsilon^2 E_2(\tau_2) & \int_{B_2}\nu_1=-2\pi i\epsilon \\
\int_{B_1}\nu_2=-2\pi i\epsilon & \int_{B_2}\nu_1=\tau_2+2\pi i\epsilon^2 E_2(\tau_1) \\
\end{array}
\right) .
\end{align}

The holomorphic quadratic differentials can be given by $\varphi_{ij}=\nu_i\nu_j$,
\begin{align}
\varphi_{11}&=
\begin{cases}
\left(1+2\epsilon^2 E_2(\tau_2)P_2(\tau_1,z)\right)dz^2 & z\in S_1 \\
 \epsilon^2 P_2(\tau_2,z)^2 dz^2 & z\in S_2 \\
\end{cases}\label{eqn:v11} \\
\varphi_{12}&=
\begin{cases}
-\epsilon P_2(\tau_1,z)dz^2 & z\in S_1 \\
-\epsilon P_2(\tau_2,z)dz^2 & z\in S_2 \\
\end{cases} \label{eqn:v12}
\end{align}
and their permutation$(1\leftrightarrow 2)$.
From the properties of the elliptic function with order two, it seems the number of zeros of $\varphi$ does not agree with the Riemann Roch theorem. Nevertheless, on $S_1$ the zero of $\varphi_{11}$ would appear near the pole since at zero $P_2(\tau_1,z)$ should become large in order $1/\epsilon^2$. Then it is expected that the zero would be inside of the excised disk and there are no zero of $\phi_{11}$ on $S_1$.Probably it is because around zeros the higher order contribution becomes significant. 

In the limit of $\epsilon=0$, $\varphi_{ij}$ becomes homogeneous or vanishes as well as the case of torus.
The existence of the pole at $z=0$, may seem a critical break down of the expansion.
Nevertheless $z=0$ is not included in the sewn Riemann surface since the small
disk around the pole is excised in order to sew together the two tori.  Furthermore also $1/\epsilon^2$ contribution of $P_2$ arises near the points sewing. Then the expansion seems to fail. 
Since the formulation of $\omega(u,v)$, however, is established in full order of $\epsilon$
we can justify the result where $|z|$ is much larger than $\epsilon$, and $P_2$ is sufficiently small there. The fact that the expansion fail around the throat, will indicate the real value of $\varphi_{ij}$ is unsuited for describing by $\epsilon$ expansion.


Now one may attempt to interpret the geometrical meaning of (\ref{eqn:v11}) and (\ref{eqn:v12}).
If the Riemann surface is affine stretched along $\varphi_{11}$, in zero-th order torus $S_1$ is deformed uniformly as known in Teichm\"{u}ller deformation of torus, and the other torus is not deformed.
In second-order both tori are deformed in complex way.
On the other hand by $\varphi_{12}$, the two tori are similarly deformed.
Since the elliptic function $P_2$ is composed of the second order pole and its summation of  images by Fuchsian group of the torus.
Near region around the throat, we may account only the contribution by the pole at the center of sewing coordinate.
Since the phase of $\varphi_{ij}$ is hard to understand geometrical role without concrete calculation, in the present status, we will only say throat is highly deformed by the blow up of the elliptic function $P_2$.

A remarkably interesting point is how the pinching parameter $\epsilon$ behaves on  the Riemann surfaces being stretched.
Nevertheless, $\epsilon$ which is one of the complex parameters of Teichm\"{u}ller space\cite{KS} should be expressed by the coefficients for $\varphi_{ij}$ and must be calculated again from $\nu_i$.
Besides, since near the throat the contribution from the elliptic function $P_2$ blow up by $1/\epsilon^2$, this description will be becoming worse for the large amount of deformation around the throat.
That would imply that the redefined pinching parameter become larger so that the expansion fails. 
For example, $\varphi_{12}\sim \epsilon$ implies the deformation by $\varphi_{12}$ largely influence to $\epsilon$.

As anticipated from these discussions, the expression may become failure around the throat of the Riemann surface. It should be emphasized that the full order expression, however, would be logically possible since the full order form of the differential forms $\omega(u,v), \nu_i, \varphi_{ij}$ is given by Yamada\cite{YA}.
Then somewhere we may see the higher order effect becomes larger than the leading order. That does not mean the breakdown of formulation rather it is not effective to describe such a peculiar part of the Riemann surface.
I also comment that other types of sewing of the Riemann surface analyzed by Yamada may be useful for analyzing such a peculiar part.

\section{Standard constant curvature metric}
\subsection{Teichm\"{u}ller deformation and dynamics of geometry in Einstein gravity}




By the ADM canonical formulation, a dynamical description of Einstein gravity becomes possible. In (2+1)-dimensional gravity, spatial two-geometry would be variables of gravitational dynamics while the diffeomorphism invariance of spacetime diminishes its number of degrees of freedom.
For example, in York's timeslicing which is a family of constant mean curvature surfaces, an  assumption of homogeneous pure gravity (possibly including cosmological constant) implies that it is consistent to geodesic slice.
Furthermore, one may simply choose vanishing shift vector. The conformal factor should be determined by solving Hamiltonian constraint and the momentum constraints force to extract a transverse traceless part of extrinsic curvature.
Then we treat an equivalence class of two-metric by conformal transformation and diffeomorphism as the degrees of freedom of gravitation.


From the viewpoint of the Riemann surface, the conformal and diffeomorphism\footnote{`Diffeo' should be assigned rather by `${\rm Diffeo}_0$', since the modular transformation is not included in this description.} class of two-metric $h_{ab}$,
is just the element of the Teichm\"{u}ller space
\[ \frac{\left\{h_{ab}\right\}}{\left\{\rm Conf.\right\}\left\{\rm Diffeo.\right\}}.\]
With the above mentioned gauge fixing, its extrinsic curvature $K_{ab}$ and two dimensional Ricci scalar $\,^{(2)}R$, satisfies the constraints for homogeneous universe which is topologically genus $g$ compact surface in pure gravity (and cosmological constant),
\begin{align}
&\widetilde{K}^{ab}_{;a}=0, \\ 
&\widetilde{K}_{ab}\widetilde{K}^{ab}-\frac{t^2}2-\,^{(2)}R+2\Lambda=0, \label{eqn:hc}
\end{align}
where $t=-K^a_a$ and $\widetilde{K}^{ab}=K^{ab}+\frac12 h^{ab}t$ is the negative trace and the trace less part of the extrinsic curvature $K^{ab}$, respectively.

Since the two-geometry is locally conformally flat, we are allowed to adopt the complex coordinate $(z,\bar{z})$ for $h_{ab}dx^a dx^b=|\Omega(z,\bar{z})|^2dz d\bar{z}$. Then the momentum constraint can be read as $\partial _{\bar{z}}\widetilde{K}_{zz}=0$ and is considered to define a holomorphic quadratic differential of $\widetilde{K}_{ab}(z)(dz)^a(dz)^b$.
As a holomorphic quadratic differentials $\varphi_{ij}\ \ (i,j=1,...,g)$ spans $3g-3$ dimensional complex vector space $A_2(R)$\cite{IT}, we might expand $\widetilde{K}_{ab}(dz)^a(dz)^b$ as $\sum_{i,j}P_{ij}\varphi_{ij}/2v$ with $v=\int d^2x\sqrt{h}$ for convenience.

Indeed, introducing the Veil-Peterson metric $G_{(ij)(kl)}$ \cite{FK,IT} into the Teichm\"{u}ller space of torus, the Hamiltonian constraint can be rewritten for vacuum pure gravity\cite{HN2},
\begin{align}
\sum G_{(ij)(kl)}P_{ij}P_{kl}-v^2t^2=0,
\end{align}
where $P_{ij}$ is not dependent on the coordinate of the spatial hypersurface.

Besides,  the Riemann Roch theorem suggests $\varphi_{ij}$ has $4g-4$ zeros.
In \cite{HN2}, the authors emphasized that the zeros imply inconsistency between these discussions and the York's slice, which means spatially constant curvature hypersurface in (2+1) dimensional pure gravity. 
The value of $\widetilde{K}_{zz}$ ($\sim$holomorphic quadratic differential) for these discussion, however, is not independent from the choice of local complex coordinate.
Actually, there always exists a local complex coordinate $z\sim \int \sqrt{\varphi}, (\forall\varphi\in A_2(R))$\cite{FG}, to make the value of the quadratic differential $\varphi$ constant though such a coordinate cannot be extended, in general, to the whole of the surface. The local coordinate is known to be related to the horizontal foliation in global. The existence of the zeros makes essential sense for this discussion as this coordinate cannot cover the whole surface prevented by the zeros of $\varphi$ as illustrated in Figure \ref{fig:horizon}.
Since the first term of (\ref{eqn:hc}) is independent of the coordinate choice, the inconsistency is not recognized globally.


These aspects may be clarified by considering a homogeneous standard metric for the Riemann surface. 
From the uniformization theorem\cite{FK}, if a two dimensional Riemannian geometry is a constant curvature, that will be realized as a fundamental region on a topologically trivial homogeneous space which is a sphere, plane and hyperboloid for positive ($g=0$), flat ($g=1$)  and negative  ($g>1$) curvature, respectively.  In other words, a Riemann surface admits any homogeneous constant curvature metric tensor. 
 For example, the conformally invariant representative $\tilde{g}_{ab}=g_{ab}/\sqrt{g}$ of a Riemann surface and such a homogeneous metric is conformally isometric.
 
On the other hand, the fact that the number of solutions of the momentum constraints is finite, means they are the global degrees of freedom of (2+1)-dimensional gravity and corresponding to the deformation of the form of the homogeneous universe.
Such a global deformation will be simply incorporated to the analysis as a  change of boundary condition of coordinates, which is the deformation of the fundamental region, for example, in an isothermal coordinate. Otherwise, one may treat it as  quasiconformal mapping of the surface, which is described by the Beltrami coefficient for its pull back of metric.

 From the Teichm\"{u}ller theorem\cite{FK,IT}  we see any metric tensor including the information of given complex structure related to the quasiconformal mapping  can be generated by the affine stretching eq.(\ref{eqn:asm}), keeping the boundary condition of the coordinates in non isothermal metric.
One may think  existence of variational holomorphic quadratic differential is inconsistent  again to such a homogeneous geometry, since for the homogeneous two-geometry the first term of eq.(\ref{eqn:hc}) is forced to be constant.
Nevertheless that does not turn out for $\widetilde{K}_{zz}$ to be constant there because it is not function but differential.

Here we will develop a homogeneous standard two-metric incorporating the information of the  Teichm\"{u}ller deformation. Since the time variation of the two-metric includes that of the holomorphic quadratic differential, the evolution of the complex structure of the Riemann surface will be caused  by the dynamics of Einstein gravity.
That will be generalized to the fact that the degrees of freedom is 
represented by the standard form of metric after fixing the gauge as follows.

By the deformation of the complex structure, the global aspects of the Riemann surface are changed.
Since the degrees of freedom will be included into the homogeneous metric as the ellipticity or `anisotropy' (which is called Beltrami coefficient by mathematicians), the treatment is similar to the Bianchi type universes\cite{WA}.
Considering the coordinate transformation on each timeslice so as to preserve the atlas, that is the coordinate mapping on the Riemann surface, pull back of the metric can include such a mobility of the Teichm\"{u}ller deformation.

By the affine stretching eq.(\ref{eqn:asm}) we have seen that the local geometry presenting the degrees of freedom of
the complex structure.
Furthermore as known in the case of torus, if the conformally invariant representative (\ref{eqn:rep}) admitted on the constant curvature surface, the metric

\begin{align}
dl^2 
=|\Omega(\zeta,\bar{\zeta})|^2
|d\zeta+k\frac{\bar{\varphi}}{|\varphi|}d\bar{\zeta}|^2
\end{align}
where $\varphi\in \langle\varphi_{ij}\rangle_{\mathbb C}=A_2(R)$ is holomorphic quadratic differential, can become that of constant curvature by the aid of conformal factor $\Omega(\zeta,\bar{\zeta})$, and then
the Hamiltonian constraint would be reduced to the independent of the spatial coordinate. 

Then the spacetime geometry is examined in a synchronous reference system $-dt^2+h_{ab}dx^a dx^b$ and the Einstein equations on the homogeneous timeslice will be given by
\begin{align}
\sum_{i,j.k,l}G'_{(ij)(kl)}P_{ij}(t)P_{kl}(t)-V(t)=0,\\
\frac{d}{dt}P_{ij}=F_{ij}[P_{kl}(t),t],
\end{align}
with $\varphi=\sum_{ij} P_{ij}\varphi_{ij}, P_{ij}\in{\mathbb C}$, where  $F_{ij}, V$ are   appropriate functionals for general relativity and $G'_{(ij)(kl)}$ are related Veil-Peterson metric.
That would  serve as the method to develop the dynamics of the Riemann surface in the
general relativity.

\subsection{constant curvature metric}
To identify the dynamical degrees of freedom, it is direct way to give a standard metric containing above mentioned global degrees of freedom of the Riemann surface.
We have seen how a complex structure determines the conformal class of two-geometry on Riemann surface, which will be treated in (2+1)-dimensional general relativity. 
Nevertheless, there is an ambiguity of Riemannian metric allowed on the Riemann surface.
We know not only one (even up to local coordinate transformation), homogeneous metric exists and it is convenient to understand its characteristics, since the complex structure separately contained into homogeneous anisotropic (non isothermal) metric and the change of global boundary condition of the complex coordinates ($\sim$atlas of local complex coordinate).
`From now on', we suppose that the boundary condition of the coordinates is trivial and static, then the dynamical degrees of freedom is not contained in it but only included into the homogeneous anisotropic metric for non isothermal coordinate as the Beltrami coefficient.	

Now we investigate the homogeneous standard metric of the double torus ($g=2$) in the leading order of the pinching parameter $\epsilon$.
First of all, we prepare 0-th order background geometry of a double torus with vanishing
pinching parameter $\epsilon$.
From the construction of the sewn Riemann surface, it can be easily found that it will be divided into two punctured tori.

\begin{figure}[hbtp]
\centering
\includegraphics[width=9cm,clip]{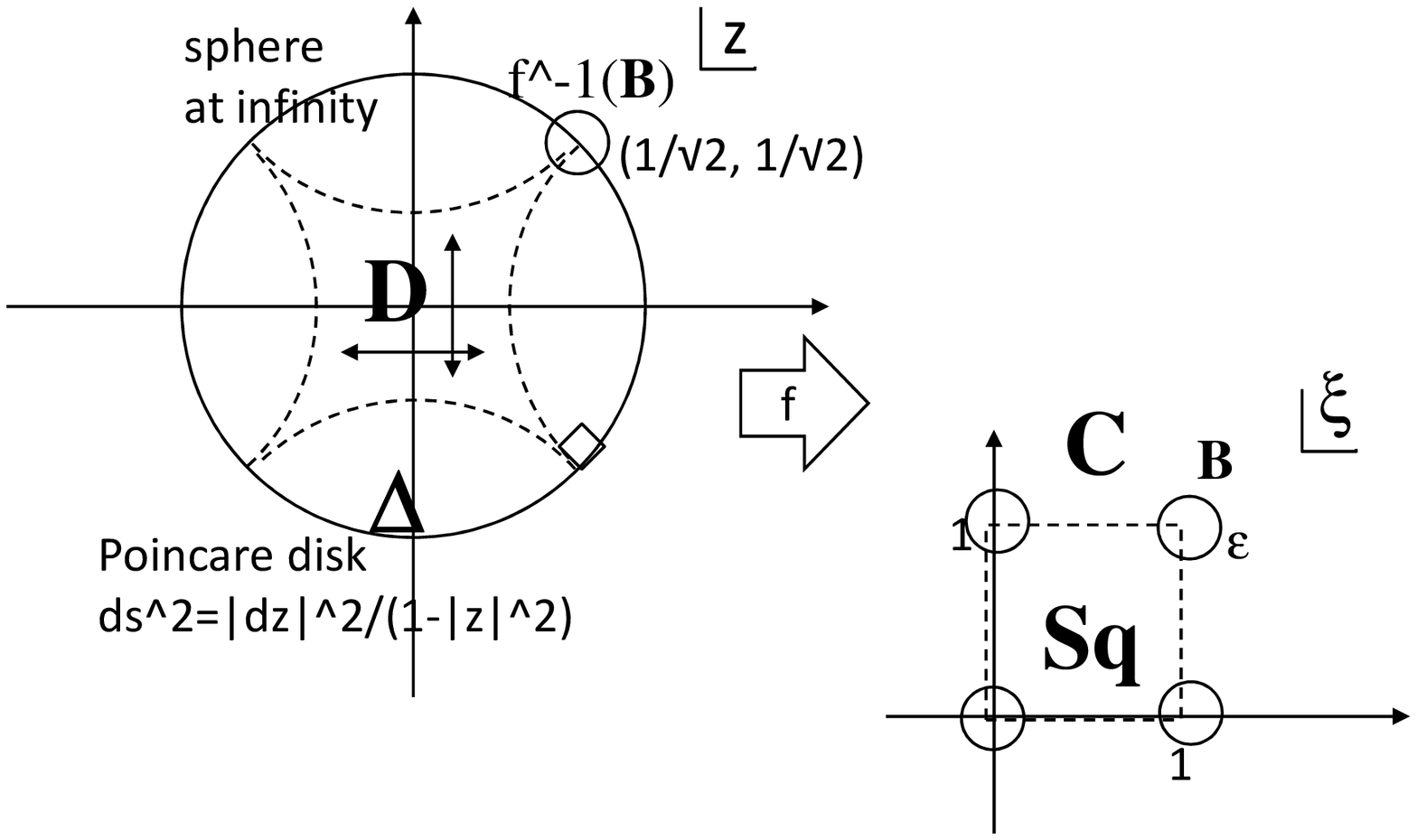}
\caption{A square ${\bf S_q}$ on $\bf C$ is to be truncated by sewing of pinching parameter $\epsilon$ and become $\bf S_q'$.
A holomorphic function $f$ on $\bf D'$ determines a conformal mapping $\xi=f(z), {\bf D'}\mapsto {\bf S_q'}=f({\bf D'})$.}
\label{fig:bgtorus}
\end{figure}

For a later convenience, we realize them with a homogeneous geometry.
Then the components become tori with a point at infinity with negative constant curvature since its Euler characteristic $\chi=0-1=-1$ is negative.
From the uniformization theorem\cite{FK,IT}, the universal cover of such a punctured torus is conformally equivalent to not $\bf C$ but $\bf \Delta$ (Poincar\'{e} disk model) or $\bf H$ upper half space.
On $\bf \Delta $ with homogeneous metric $g_P=\frac{4dz d\bar{z}}{(1-z\bar{z})^2}$, the fundamental region $\bf D$ of the punctured torus is a region surrounded by four circles passing through $z=\pm \frac1{\sqrt{2}}\pm i\frac1{\sqrt{2}}$. Since they are perpendicular to the sphere at infinity ($|z|=1$), they are geodesics crossing at the infinity, (see fig. \ref{fig:bgtorus}). Opposed edges are identified to form a torus and then it carries a complex structure of a punctured
 torus. 

Though we should obtain a unique bilinear two form there, it is convenient to consider  `conformal mapping' $\xi=f(z)$ by which the region $\bf D\subset \Delta$ is mapped to a region ${\bf S_q}=\{0\leq \Re\xi\leq 1, 0\leq \Im \xi\leq 1\}\subset \bf C$.
Here we should be careful that  a disk $\bf B$ in $\bf C$, with radii $|\epsilon|$ and  $f^{-1}({\bf B})$in $\bf \Delta$,  will be excised in sewing process.
Rigorously, a truncated $\bf D'$ is conformally mapped to ${\bf S_q'}=f({\bf D'})$, though $|z|=1$ implies $f$ is not holomorphic, there.

In the coordinate $\xi$, pull back  $f^*g_P$ of metric $g_P$ is given, and the quadratic differential is determined along the analysis in previous section. Then the affine stretched metric is given by

\begin{align}
ds^2=\frac4{(1-z\bar{z})^2}\left|\frac{\partial z}{\partial\xi}\right|^2\left|d\xi+k\frac{\bar{\varphi}}{|\varphi|}d\bar{\xi}\right|^2 \label{eqn:asm3}\\
\varphi=\sum_{i,j}Q_{ij}(t)\varphi_{ij}(\xi)\\
Q_{ij}(t)\in {\mathbb C}, \varphi_{ij}\in A_2(R)
\end{align}

Of course, inhomogeneous $\varphi$ implies the metric is inhomogeneous.
In order to compensate the inhomogeneity, we consider another conformal transformation of the metric by $\tilde{g}_{ab}=|\Omega(\xi,\bar{\xi})|^2g_{ab}$ where $\Omega(\xi,\bar{\xi})$ should satisfy the homogeneity equation (c.f., in $0$-th order $\varphi$ is homogeneous). 
 As $z(\xi)$ is the bi-holomorphic transformation, $\Delta_{\xi,\bar{\xi}}\ln \frac{\partial z}{\partial \xi}=0$, and $|\varphi||d\xi +k\frac{\bar{\varphi}}{|\varphi|}d\bar{\xi}|^2$ is pull back of $dz d\bar{z}$, it turn out that $|\Omega|^2=|\varphi|$ makes  (\ref{eqn:asm3}) constant curvature, 
     \begin{align}
\tilde{ds}^2=\frac4{(1-z\bar{z})^2}\left|\frac{\partial z}{\partial\xi}\right|^2|\varphi|\left|d\xi+k\frac{\bar{\varphi}}{|\varphi|}d\bar{\xi}\right|^2 .\label{eqn:shm}
\end{align} 
 

That would be a standard metric as another representative of an equivalent class of metric by conformal transformation and diffeomorphism of coordinate transformation.

Of course, one can easily confirm that the metric satisfy momentum constraint by checking transverse traceless component of extrinsic curvature will be proportional to the holomorphic quadratic differential $\varphi$, since that is covariant under spatial coordinate transformation corresponding to the affine stretching.

Here it should be noted that there is a `gauge freedom' in conformal mapping, since
holomorphic function $f$ s.t. $f({\bf D'})={\bf S_q'}$ allows still arbitrariness.
While different holomorphic function $f'$ gives another pull back $f'^*g_P$ of metric $g_P$ which is conformally equivalent to $f^*g_P$, the determined quadratic differentials are same.
Consequently the resultant standard metric is equivalent to that of $f$ up to coordinate transformation  since the uniformization theorem indicates a Riemann surface admits a unique diffeomorphism equivalent class of a homogeneous geometry.
The gauge would be fixed for any convenience of calculation or we will examine any gauge invariant variables like a transverse traceless component of the extrinsic curvature, which will be independent of `$f$' except for its condition supposed here.

Then we can set $3g-3$ variables $\sim Q_{ij}$ and its time derivatives as initial values.
Consequently, we see there are $6g-6$ degrees of freedom  contained in eq.(\ref{eqn:shm}).



\section{summary and discussions}
Dynamical degrees of freedom of a higher genus Riemann surface universe in (2+1)-dimensional gravity has been investigated.
A homogeneous standard metric that is a representative of equivalent class by conformal transformation and diffeomorphism of Riemannian metrics on two dimensional surface, have been determined in the leading order of the pinching parameter.
Then we see that $6g-6$ degrees of freedom is contained in the standard metric.
Especially for $g=2$ Riemann surface, that is so called double torus, we have made a concrete calculation which provides that up to diffeomorphism.
Such a homogeneous metric will make it possible to formulate the dynamics of Teichm\"{u}ller deformation in the context of ADM-formalism of pure $(2+1)$ dimensional general relativity.
Though only for double torus we demonstrated to have such a homogeneous standard metric, extension to general Riemann surfaces is straightforward since they are decomposed into tori to be sewn in pinching parameters.

Furthermore though we have carried out the analysis only in the leading order of the pinching parameter, a similar analysis will be possible since in Yamada's analysis\cite{YA} the differential structure of such a sewn Riemann surface has been given in the expanding series of the pinching parameter.

As shown in the present article, the result is in approximation under the small pinching parameter $\epsilon$.
 The existence of the error would indicate that the momentum constraint is not exactly satisfied. That implies the  present result is not exactly covariant under coordinate transformation. Then further investigation about the higher-order analysis will reveals correct bi-holomorphic aspects of complex structure of sewn Riemann surface by the pinching parameter.

In our forthcoming work\cite{FW}, the whole of the Einstein equations will be solved and reformulated for canonical formulation.
Moreover the Teichm\"{u}ller deformation along the holomorphic quadratic differential should be visualized, for example, in the relation with the horizontal (or vertical) foliation.
It requires organized analysis and consideration by concrete complex analysis concerned with the elliptic functions. The result would be reformulated independent of gauge freedom `f' for convenience, for example transverse traceless part of extrinsic curvature, or any variable in Teichm\"{u}ller space.
Especially, the relation between the variables introduced in this analysis and the periodic matrix which is usually used in formulation of conformal field theory in the higher genus Riemann surface.

In a range of the present investigations, we have never concerned the modular transformation of the Riemann surface, while that is mathematically well established \cite{FK} as well as that of torus is well known.
For example, on considering quantum theory, that will be very important\cite{HN2}.
We will be also approach that in detail in forthcoming study, and that will be helpful for investigation about quantum summation of topological degrees of freedom\cite{MW}\cite{KS}.

From analytical reasons, the elliptic function plays essential role in the present investigation. On the other hand, some elegant algebraic theory for the elliptic function, e.g, imaginary multiplication, have known and established.
Through studies of the gravitational theory of the Riemann surface, one may expect that such a mathematical knowledge provides entirely new progresses of especially quantum theory of gravitation.

\section*{Acknowledgements}
We would like to thank to Prof. H. Miyaji for his helpful advices about mathematics of Riemann surface.


\end{document}